# Teaching and learning physics: A model for coordinating physics instruction, outreach, and research

Noah D. Finkelstein[1]

*Abstract. This paper describes the development of a new university physics course designed to integrate physics, education, research, and community partnerships. The coordinated system of activities links the new course to local community efforts in pre-college education, university education, university outreach, and research on teaching and learning. As documented both by gains on conceptual surveys and by qualitative analyses of field-notes and audiotapes of class, the course facilitates student learning of physics, as well as student mastery of theories and practices of teaching and learning physics. Simultaneously, the course supports university efforts in community outreach and creates a rich environment for education research. The following narrative describes the motivation, structure, implementation, effectiveness, and potential for extending and sustaining this alternative model for university level science education.*

**I. Introduction.**

The explicit mission of many large-scale research universities includes three core elements: the pursuit of excellence in research, teaching, and community service. However since the mid-twentieth century, many universities have heavily emphasized research without equal commitment to teaching or community service. Efforts directed at supporting high quality teaching (at the university or pre-college level) and partnerships with the communities that house the universities are largely treated as separate, and often non-essential, programs at these institutions of higher education. This paper addresses the question of how such institutions might begin to coordinate these three seemingly disparate elements of the university mission into a single activity system that enhances all three.

The focal point of the coordinated system is a class entitled *Teaching and Learning Physics* offered within the physics department (Finkelstein, 2003). The emphasis of the present work is to describe the structure of the class and the impact of the class on students. Through the use of pre- and post- tests of students' conceptual grasp of the physics content, audio tapes of classes, ethnographic field-notes, course evaluations and student interviews, the class is presented as a case study to demonstrate that such an approach is useful for improving students' grasp of physics and of teaching. At the same time, in addition to documenting the effects that the course has on students, this study examines how well this environment is suited for physics education research, how supporting and surrounding institutions respond, and the potential for sustaining such a pursuit. The discussions of lines of educational research, and the likelihood of sustaining this course follow the course description and student evaluation.

[1] Department of Physics University of Colorado, Boulder. The author may be reached at noah.finkelstein@colorado.edu © Noah Finkelstein, 2004 all rights reserved, do not quote without permission of the author. This research was conducted with the support of the National Science Foundation's Post-doctoral Fellowships in Mathematics, Science, Engineering, and Technology Education (NSF's PFSMETE Grant Number: DGE-9809496) under the mentorship of Michael Cole (University of California, San Diego) and Andrea diSessa (University of California, Berkeley). I wish to thank these mentors and my colleagues at the Laboratory of Comparative Human Cognition and those in the Department of Physics (Barbara Jones, Edward Price and Omar Clay) at UCSD for intriguing discussions, insights, and their support. Finally, I am grateful for the support and critical feedback of the members of the Physics Education Research Group at the University of Colorado.

The effort to create a coordinated system of teaching, research, and community partnership builds on recent efforts to support student learning in physics, to incorporate education research within departments of physics, and to address a critical shortage of teachers and lack of diversity at the university level. More and more widely, university faculty acknowledge that the traditional lecture-style physics course fails to impart a deep-seated conceptual understanding of course content (Hake, 1998; McDermott and Redish, 1999; Redish, 2003). As a result, in some institutions, a new breed of physics class is evolving -- one that encourages student engagement. Coupled with this recognition, the physics community is beginning to re-assess both the goals of undergraduate courses and what constitutes the discipline more broadly. One outcome of this reassessment is the idea that education research is an integral part of the discipline of physics (APS, 1999). Another outcome is that more departments of physics and schools of education acknowledge the need to better prepare teachers of physics (Schmidt et al. 1999; TIMSS, 1999). Furthermore, in California and elsewhere, a host of political initiatives and educational reforms have challenged the University's ability to meet its charter commitment to serve all of the state's population.[2] At the same time, studies of service learning programs, those that send university students to engage in community-based activities as part of their education, demonstrate significant and improved outcomes for students engaging in these activities (Astin et al, 2000; SLCH, 2003). As a result, a significant response from both the legislature and the university system is to support community outreach in an effort to better prepare current and potential students, especially those from traditionally under-represented populations.

This research program addresses these related problems: 1) the improvement of student interest, understanding, and expertise in physics, teaching, and learning; 2) the creation of community-based activities which address the outreach and service interests of the university; 3) the provision of a research site for the study of the teaching and learning processes. The coordinated ensemble, represented by this course, is an opportunity to merge these many agenda. Such an effort follows the work of Cole and others who create rich, theoretically motivated environments that foster student learning and support fundamental research on development and culture (Cole, 1996; Cole, 1998).

**II. The Activity.**

The course on Teaching and Learning Physics is composed of three elements: a study of physics content, readings about the teaching and learning of physics, and practical experience teaching physics to less educated students. The course is designed for upper-division undergraduate physics majors who have expressed an interest in education. It is described in the course catalog as:

> A course on how people learn and understand key concepts in introductory physics. Readings in physics and cognitive science plus fieldwork teaching and evaluating pre-college students. Useful for students interested in teaching science at any level. Pre-requisites: [introductory level courses in electricity and magnetism]

---

[2] The anti-affirmative action debates have received widespread publicity and response both at the state level and at the University of California level. The passing of California Proposition 209 in 1996 was the culmination of many of these debates.

Each of the three curricular components of the course (physics content, theories of teaching and learning, and practical teaching experience) represents roughly one third of the course. One of the two weekly class sessions focuses predominantly on the study of traditional physics content. The other emphasizes readings in physics education and cognitive theories of learning. At least once per week, students engage in the laboratory portion of the course, teaching college and pre-college students.

Each of the course components is designed to complement the others by explicitly providing varied perspectives from which to view physics. Because the course draws upon and addresses questions from different domains (physics, education research, and community outreach), it sits at the interfaces between each of these domains and borrows material and methods from each of these bordering worlds (Star, 1989).

Not only do the course participants benefit from the variety of resources, but also by acting at the interfaces of disciplines, the class provides a mechanism for communication between, and coordination of, these differing domains. For the department of physics and the teacher education program, the course serves both as a catalyst for improving the university students' conceptual understanding and as a common object of discussion and coordination for departments. For physics students, the class acts as an amplifier and reorganizing mechanism for their physics knowledge and as portal from physics into education and teaching. For the outreach program, the course strongly links university efforts in science to community-based education of children. Figure 1 illustrates some of these relations. The figure depicts the three interacting components of the course as the vertices of a triangle. Each of these components necessarily interacts with and in fact co-constructs the others, as will be described in detail below. As a discipline, physics addresses content and the teaching of content. Education concerns itself with the theories and practice of both teaching and learning. Lastly, efforts in community outreach blend the practice of teaching (fieldwork) with content (physics) in community-based settings. Of course, the boundaries of these domains and activities are not fixed, nor are they mutually exclusive.

A program that brings together a study of science content, study of educational and teaching theories, and practical experience teaching science content is remarkably rare, if not unique. Usually, these components are separated. For example, in education schools there are science teaching methods classes, where there is some blending of content and pedagogy. In various science departments, there are an increasing number of classes on cognition and student learning. Also, in various portions of university, there are an increasing number of service-learning or practicum classes where students are guided in teaching experiences. However, each of these approaches differs from design and mission of Teaching and Learning Physics, which strives to blend all three elements.

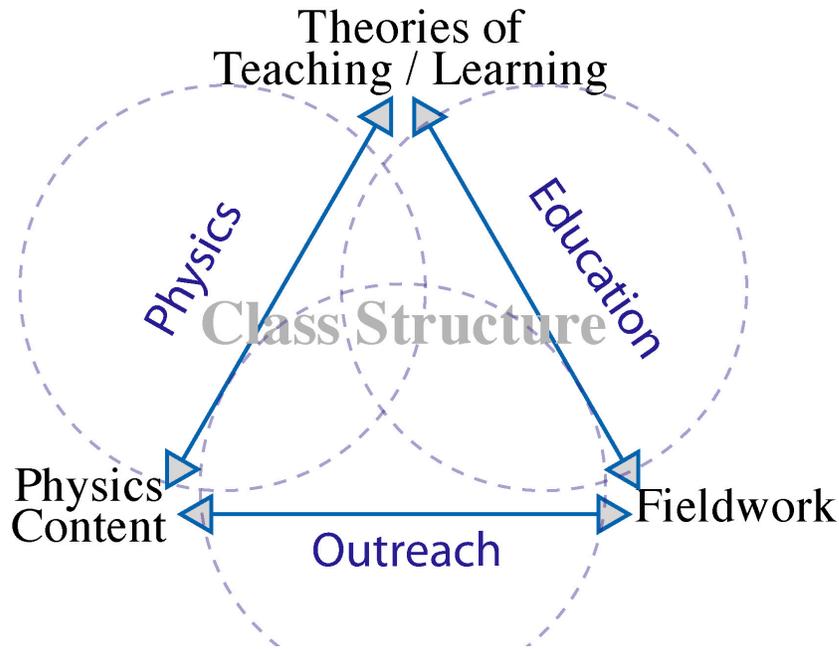

Figure 1: Course structure and disciplinary boundaries

In this model for a physics course, the students engage in activities that engender both broad-based skills which span these domains (e.g. problem solving, analysis, and meta-cognition) as well as specialized domain-specific knowledge and skills (e.g. physics content, and knowledge of and practice in theories of teaching). In addition, the course is designed to be flexible enough to capitalize on the emergent nature of the activity. That is, because the participants, locales, and even the content are dynamic in nature, the precise form of the activity changes over time. The assertion is not that physics itself is changing (though many may argue about the social construction of the discipline), but rather since the course structure is flexible, it allows the coordinated activities to adapt to local context. The arrangement of the components of this activity system (the vertices in Figure 1) may be thought to be skeletal in nature, and the actual content, interaction, and environment form the "flesh" that is placed upon the structure.

**III. The Organizational Details.**

The course is designed for upper division students who have covered at least a minimum level of lower-division coursework in physics. The class meets three hours per week on campus (covering traditional physics content and theories of student learning), and engages students in two to four hours per week of practical experience, teaching in local community centers and schools. Each component of the course focuses on the domain of electricity and magnetism (E&M). As much as possible, each component is integrated with the others; the lines between the activities are purposely blurred. A student reading about theoretical difficulties in understanding the concept of electric field is encouraged to wrestle with his own understanding of the topic. Furthermore, as much as possible, there is a temporal alignment of the activities. The same week that students study electric fields, they read about student difficulties in understanding the concept of fields, and also attempt to teach the concept to others.

The first component of the course covers traditional physics content, approximately two-thirds of an introductory course in E&M (using texts such as Halliday, Resnick and Walker, 1997). While calculus is used in the analysis of problems, mathematics and symbol manipulation are not emphasized. Rather, each topic is introduced from a conceptual viewpoint, and placed within a broader context of other topics in physics. Similarly, the physics segment shifts in focus from symbolic representation and the coverage of text to an active engagement of the students in project-oriented lessons, which foster active construction of models of physics. The class follows a constructivist approach. That is, the course emphasizes learning as a personal and social act where the learner actively builds up understanding using local resources, rather than passively accepting knowledge that is transmitted from the teacher to the student. Most often these lessons focus on the physical construction of material and public presentation. The added level of emphasis physical construction and public display places the present approach within what Papert calls the constructionist camp (Papert, 1991). These course activities vary from Tutorials (McDermott et al., 2002), to discussion, to group problem solving (Brown and Campione, 1990; Brown, 1992), to teaching and materials development. The lessons are designed to force students to confront traditional difficulties within electricity and magnetism (Posner 1982). The class encourages student learning during class hours, rather than solely after hours. Homework is assigned, but emphasizes the conceptual understanding of content. For example, for traditional textbook-based problems, students reflect on the solution process and critique the problem, in addition to deriving an answer. Other homework assignments include interviewing or teaching novices about advanced concepts in E&M and subsequently writing-up the process and results. Each of these practices is designed to foster development in two domains: mastery of content and improvement of meta-cognitive skills, i.e. reflection, regulation, and epistemological development (Schoenfeld, 1986).

The second course component, readings in physics education research, occurs in a seminar format. Each session begins with brief student presentations followed by discussion. Students support or refute ideas presented in the readings using evidence from the other components of the course. Readings in physics education research fall into several categories: empirical research on learning (McDermott and Shaffer, 1992), theoretical underpinnings of learning physics (diSessa, 1988), and cognitive science approaches to teaching and learning processes more generally (Brown et al., 1989). Students hand in weekly notes with summaries or questions relating to the readings. The notes are commented upon and returned to the students. These informal notes insure that students read the assigned papers, and force some level of reflective analysis.

Student teaching occurs at one of four sites, in and after school hours at the junior and senior high school level. Students are encouraged to develop and teach their own curriculum (within E&M); in each instance, supervisors, both at the university level and at a local level in the partnering junior or senior high school programs, oversee student work. In this fashion, student fieldwork differs from many traditional service-learning models, as the students are guided and are studying both the content and the practice of teaching while engaged in the process itself. Each week students and supervisors write detailed field-notes describing their experiences, curriculum, interactions, and reflections. In addition to using their experiences at the field-sites as a proving ground to test and refine theories of education, students use these sites as resources for research for final projects and papers. Again, the final papers are a mechanism for students to reflect back upon the quarter's activities. Though not a goal made explicit to the students, this teaching experience is designed to help students master physics content as well.

**IV. Evaluation and Discussion.**

Course evaluation occurs at several levels: at the level of student learning, as a research venue, and as an organizing tool for institutional coordination enabling outreach. The data are presented as a proof-of-concept to demonstrate that this class has the potential to improve student understanding of physics and teaching and learning principles, to serve as a rich venue for research, to provide an avenue for community partnership, and finally, to coordinate these activities into a cohesive whole where the individual components complement one another. The present work predominantly focuses on evaluation at the student level. However, no less significant is the analysis of this system as a research site, or the role that this activity serves in the coordination of various institutions. The data presented are primarily from the first offering of the 10-week (one quarter) course, which enrolled 14 students at a large research university in California.

*A. Teaching / Learning -- student expertise in physics*

Students' improved capabilities in the domain of physics were of primary interest. It is worthy of note, however, that students generally did not enroll in this course to remediate their understanding of physics. All students in the course had passed one, two or in some cases, three classes in electricity and magnetism. Nonetheless, all students demonstrated improved understanding of the domain. Evaluation of student performance included: pre- and post- test of basic concepts in electricity and magnetism (described in more detail below), audio-recordings of class sessions, student evaluations of the course, and in-class observations. All students who completed the course (N=13) participated in all forms of evaluation with the exception of days when students were absent from class.

The diagnostic test was a mix of thirty-five free-response and multiple-choice questions drawn from the Conceptual Survey of Electricity and Magnetism (Hieggelke et al., 2001), the Electrical Circuit Concept Evaluation (Sokoloff et al., 1998), and two original questions.[3] In addition to selecting answers for each question, students provided confidence levels for their answers on a 3-point Likert-like scale (guessing, somewhat sure, certain). Results of the pre- and post- test are shown in Figure 2. The independent axis of the plot lists individual students. The left most student, A, had never formally studied the material and withdrew from the course. The right most student, N, was a fifth year graduate student in physics. A dashed line indicates a division between physics majors and non-majors. The dependent axis plots student performance. The mean pre- and post-test scores are respectively 54% ($\sigma$= 25%) and 74% ($\sigma$= 24%). The average of individual student gains is 51% ($\sigma$=30%; N=13; p<0.001).[4]

---

[3] The following questions were used from the CSEM: 1-2,6-8,10-11, 15,17-18,20-22,25
The following questions were used from ECCE: 1-5,9-10,12-14,27-32
Additional questions were developed to augment these instruments with free response questions of the same content, and to cover concept of flux. These added questions are similar to the conceptual questions presented in Halliday, Resnick, and Walker (1997).

[4] The measured gain is similar to the gain Hake (1998) reports : (post-pre)/(100 - pre). However, here, the average of individual gains is reported rather than the gain of the class average. Inverting the order of operations (averaging and measuring gain), shifts the statistical weighting of individual students. The measure of significance is evaluated by means of a single sample t-test of the gain defined above.

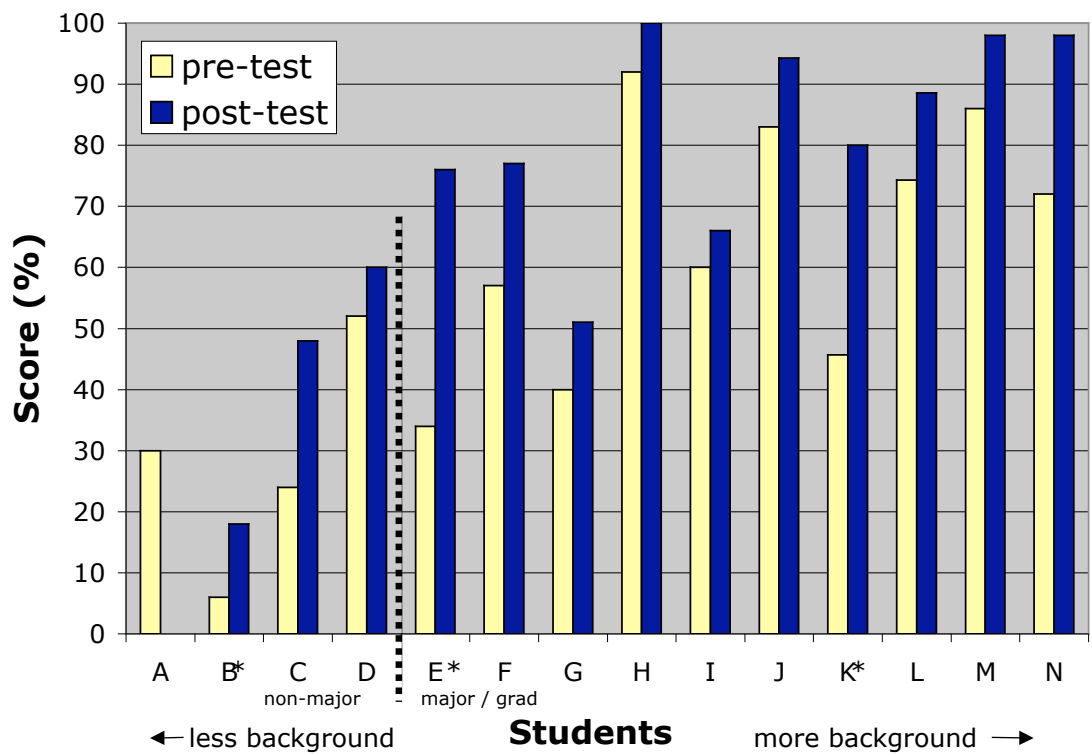

Figure 2: Student pre and post-test scores on conceptual survey of the basic ideas in electricity and magnetism. Mean pre- and post-test scores are respectively 54% ($\sigma$= 25%) and 74% ($\sigma$= 24%). The average of individual student gains is 51% ($\sigma$=30%; N=13; p<0.001)

    Aside from demonstrating improved conceptual understanding of physics, a few points are worthy of note. No student had a complete mastery of the *most basic* concepts at the beginning of the course, despite each student having had some background in physics (and having covered this same material previously). While it is not argued that this class is the most effective mechanism for students to learn concepts of physics, it is clear that it is an effective mechanism for increasing student mastery of basic concepts. Furthermore, upon entering the class, some of the students, even a physics major (Student E), performed at levels roughly equivalent to the un-schooled student, Student A. Generally, those students who had a better grasp of the material upon entering this class made greater improvements than those who were weaker at entry. (The half of the class that performed best on the pre-test made average gains of 66%; whereas, the bottom half of the class made gains of 31%.) Perhaps this is due to the challenge of offering a class to such a diverse range of students.[5] The students' backgrounds spanned a range of eight

---

5    The class was designed for students who had some familiarity with the material at the outset. As a result, the course was better suited for those students who performed better on the pre-test. However, this is not to say the class model could not be used for an introductory level, or for the lower performing students, but rather the class could not equally well address all of the students who spanned a range of 8 years exposure to formal physics.

years of exposure to physics. The * next to the student letters on the bar graph indicates female students. On average, there was no difference in performance by gender. However, the two greatest improvements in absolute score (post-test less pre-test score) were both women (Students E & K). In this case, there is some suggestion that while there may be some correlation between gender and class performance (Students E & K show 63.5% gains), it is masked by familiarity with the material (by including Student B, the average gain drops to the class mean).

The audiotapes of classes and observational notes written immediately following each class serve as complementary tools for evaluating student understanding in a qualitative fashion. These ethnographic observations are full of examples that corroborate the pre-test data -- students do not begin the course with the expected grasp of material. For example:

> From this discussion it became very clear that [Student C], whom I had asked to step to the board, didn't really understand electric fields all that well (the topic had recently been covered in [this course], and the class pre-req.) ... it was clear that the discussion helped 2 people in the room [Students B and C], was probably useful for [Student I] (whom I often caught guessing).- Day 12

Similarly, class discussions reveal when students may not have a thorough grasp of the material. In a reporting on a research study of students' difficulties with elementary circuits (McDermott and Shaffer 1992), Student F reveals some of his own difficulties on audiotape, stating:

> *Student F:* The point is: students tend to reason sequentially and locally rather than holistically. The students don't really see the big picture. ... if you take one light bulb out of the circuit, what would happen? And if it is in parallel or in series is there going to be a change? and the students are not understanding that.
>
> And that's actually where I had ... an interesting thing ... like one of those things we could go over is ummm if it's ...they're doing one of those in series umm ... they're talking about the switch, and Figure 5 [student reads:] "since the total resistance of the circuit would increase, the current through bulb A would decrease, and it would be dimmer." And to me, my gut feeling say that it would become brighter, which is kinda interesting. - Day 9

The use of audiotape and notes helps detail when and why students make conceptual shifts. For example, in a discussion about one of the course readings on the use of analogies for teaching electric circuits (Gentner and Gentner, 1983), a student reflects on the utility of a water reservoir analogy:

> *Student F:*   Can we just talk about…like if you have the uhhh two batteries in series. You get twice the current
>
> *Instructor:*   right

> *F:* Which actually [pause] taught me something. I always thought the batteries in parallel gave you [inaudible]
>
> [discussion of the water analogy and two batteries in series produce twice the current of a single battery for a fixed load]
>
> *F:* okay but see, I thought it was the opposite of that. Because I think I was using the wrong model ...
>
> [Audio tape transcription of class Day 7]

From my notes written immediately following the class:

> Student F made a very interesting revelation, with which Student J also identified: Total lack of conceptual understanding of series and parallel batteries and bulbs. Student F made the comment that he had thought batteries worked differently until he had read the article. Still, throughout the class he and Student J would make little mistakes about relative brightness, voltage etc. When asked to think about it in terms of the article and the analogies presented, [however,] they would get the right answer. It required conscious effort and thought. - Day 7

A comparison of the pre- and post-test responses for students F and J confirms both that the students better understood how elements behave in series and parallel and that the students had greater confidence in their answers. On the series and parallel circuit questions in the conceptual survey, Student F improves 33% from pre-test to post-test (changing from 70% to 80% correct) and his confidence in his answers rises from 1.8 to 1.1 (where 1 is certain and 3 a guess). Student J improves 75% (60% to 90%) with confidence rising from 1.7 to 1.1.

Of course, not all student comprehension increases linearly. These notes and class recordings capture instances when students' understanding regresses and detail the conditions that lead to the retreat in understanding. One particularly interesting case details the retreat in understanding of Student D who changes models for current flow in a circuit. On the pre-test, the student consistently demonstrates a more expert view. On the post-test, the student consistently uses a more naive model of current consumption. Analysis suggests that this student learned of the more naive model from class discussion, and in particular from another student. Such findings are consistent with others who argue that a more confident, but less expert student may convince a more advanced but less confident student to adopt the more naive view (Hogan and Tudge, 1999). These notes allow for in depth case-studies of students which provide insights into the mechanism behind student achievement, or lack thereof.

Understanding and learning physics is intertwined with students' attitudes. By quarter's end, in an open ended 'comments' section of the course evaluation, students report on their own understanding of the material, and their greater comfort and interest in the subject area:

> I'm finally enjoying this material [E/M ...] Overall, I've
> learned (understand finally) so much about E & M and I'm
> learning about techniques to teach it - week 5
>
> I learned a lot about teaching, and even found a new
> interest in the subject of physics through this course - week
> 9
>
> [The best part of the class was] discovering that I didn't
> know what I thought I knew about physics - week 10
>
> I'm not good at [discussion]. This is really the first class
> where I have really had to talk about what I think- week 10

The goal for students in this course was not simply to improve their conceptual understanding of and attitude towards physics, but also their epistemological development (what it means to know physics) and their awareness of their own understanding. Following Hammer's metric of epistemological development in physics (Hammer, 1994), there is some suggestion in these data that students are moving from a belief that physics is a mastery of disjointed formula handed down by authority to a belief that physics is a coherently organized and related set of principles useful for an independently developed understanding of the world. Furthermore, it is suggested in the above quotes (and those reported later in the paper) that students become more aware of their own knowledge. For example, in a discussion about current conservation, Student F reveals:

> I don't know some of these things. I have the same
> misconceptions that kids and undergraduates that we're
> reading about. I'm a physics major, and I don't know these
> things. I can do the advanced stuff (calculations etc...) but
> not the conceptual side. - Day 9

and field-notes document a similar response by another student:

> Student J detailed his experience of not believing in current
> conservation. He also identified where this belief arises
> from. Ironically, such thought hasn't been countered by
> any formal training. He was a little [upset] about this. -
> Day 9

In terms of Schoenfeld's definition of metacognition, students are self-assessing, which is a necessary precursor to regulating their knowledge of physics (Schoenfeld, 1986).

To summarize briefly, students develop greater expertise in physics broadly conceived. Students demonstrate gains in conceptual mastery, attitudes, beliefs of what constitutes physics, as well as their ability to monitor and potentially modify their own level of understanding of physics.

*B. Teaching/Learning -- student expertise in teaching*

The structure of the course was motivated by the belief that such expertise is strongly influenced by students' experiences teaching. In line with this hypothesis, students report improved ability and interest in teaching. In the 'comments' section of course evaluations students report:

> I got so excited [about teaching]-week 10
>
> I thought I had a pretty good grasp on how to teach physics, but I've learned enough to revamp my whole style-week 9
>
> I loved fieldwork b/c I actually was able to observe the teaching theories involved in class and even put them into practice-week 10
>
> This [fieldwork] really drove home some of the points made in our discussions and readings-week 10

Students also report that their conceptions of what constitute teaching changed. During the first and last weeks of class, students turned in "statements of teaching," where they were charged with writing a paragraph or two on their approach to teaching and teaching philosophy. They reflect upon what it means to teach:[6]

> <u>Student L, Pre</u>: ... there seems to be two ways of going about [getting people to learn]. One school of thought is that repetition is how one learns, and the teacher should focus on the most important ideas and go over them repeatedly. The other methods is to saturate the students with information... I have no opinion on which method works better...
>
> <u>Student L, Post:</u> I believe that teaching is less telling and more leading through interactive experiences. It is important for a teacher to know the subject material and be able to convey it clearly, but it is equally important for a teacher to be able to prompt students into learning experiences through which students learn on their own, and in the process own the knowledge themselves. ...Another important duty of a teacher is to provide an environment for the student that is conducive to learning. This may include ... providing groups of students for interaction and making sure the students are learning and not just memorizing by getting involved in the learning process.
>
> <u>Student E, Pre</u>: I think that the most important thing to do when teaching physics is to keep the class's attention. This

---

[6] Though only three statements are presented here, these responses are representative samples, rather than extra-ordinary student statements.

> can be done by inspiring students ... making physics ... relevant to their lives, by being humorous or animated ... Make physics class an inviting atmosphere and hold class discussions.
>
> *Student E, Post:* My teaching strategy this quarter in class and at site has focused on creating a solid foundation of physics concepts for the students through hands on activities ... I've made a conscious effort ... not to make previous assumptions about one's knowledge ... I think that group work and project based learning is a more successful way to go than just lecturing
>
> *Student H, Post:* I have gained invaluable experience in (and learned the main underlying principles of) teaching, both in general, and as it relates to physics. I think this experience has helped me to refine my goals, strategies, and implementation for teaching. ... I also was able to see just how important it is to keep students actively involved with the lesson, participating in through-provoking projects, thinking, answering questions, asking questions, explaining, and discussing ... These activities are where the real learning takes place, not half sleeping through a lecture on the finer points of proving the Schrödinger equation

It should be clear that the class holds a heavily constructivist bent (Papert, 1991), which seems to have seeped its way into the students' consciousness. While may be argued that students were parroting discussions from class rather than shifting epistemological and pedagogical view-points, evidence from students' discussions in class, their field-notes, and final papers, suggests that the students constructed a rich framework of inter-related ideas about teaching and learning. A significant effort was made to ensure that students wrestled with the theoretical underpinnings of their convictions and teaching experiences. Some of these theories and tools for understanding the teaching / learning process begin to cycle through public communication in the course as demonstrated by an increased use of technical language from the course readings in student field-notes. For example, Student H writes of pre-college students' failure to grasp a lesson, "This might be a consequence of the fact that they were not forced to confront many of their pre-conceptions, come upon a conflict, and resolve it." These sentiments parallel Posner's comments on developing a theory of accommodation (Posner, 1982). The field-note continues, "knowledge ... never really became integrated as a system," which, in this context, appears to refer to diSessa's notion of knowledge in pieces (diSessa, 1988) and Reif's discussion of knowledge structures (Reif, 1986). Students adopt strategies from the readings and reflect on their own success and failure to implement these strategies in the teaching environment.

Based on observations, students' field-notes, and student final projects, there is strong indication that students became better teachers. Students were found to implement and evaluate practices discussed in class, to research other methods of teaching, and to appropriate these for use in their own teaching environments. Students constantly evaluated their own practices (and each other's). For example, in Student F's field-notes, he reflects on the effectiveness of two approaches to teaching:

> [The high school] students seemed to respond fairly well to the light bulb/ resistor box experiment, but seemed bored when explained to them by theory on the white board. After the explanation, many students were not able to guess [correctly] about the change in brightness of the bulb as the resistance in the series with the bulbs changed. Only after they were able to play with this themselves, were the students able to make theories. - Week 7

Approximately half of the final student projects were directed at assessing pre-college student performance and how performance correlated with such variables as teaching style, learning environment, representational form of the material, or gender. These studies served to confirm or refute others' theories of student learning, and to evaluate which strategies work best for the University students in their working environments. For example, in a study of the effectiveness of representational forms (white-board versus worksheets), Student L confirms the benefits of active-engagement and begins to examine why this works in his classrooms:

> To see why these two environments [high school and college] yielded such opposite results, one must contrast the situations of the students involved. One can expect that when students learn from a lecture format lesson, they will not be able to apply the concepts as abstractly as when they were involved in the learning. Not only will the students be merely watching and not participating, but also it is quite likely that hey will not keep interest in the presentation … At the [college] session, the students were actively learning, discussing and sharing. The [high school students] were instructed to draw diagrams, whereas the U[niveristy] students were using diagrams as tools to reach a goal – finding a solution to a problem. - Final paper, Student L

Student L uses the opportunity of teaching and of conducting a study of his students' learning to develop his own theories of and strategies for teaching.

In summary, because of the coordinated activities of the course, students demonstrate a greater grasp of both physics and of teaching, and student improvement in each of these areas is broad and multi-faceted. Students demonstrate an improved grasp of content and application of content.*C. Potential for Research*

This course provides a valuable research venue for making insights into the process of learning physics. While a host of such opportunities exist, the present focus emphasizes undergraduate learning. Data from student work, interviews, audio taped classes, and field-notes suggest how the course affords insight into the importance of context in the learning process. The sections above suggest several features of the context and content of this program -- the interplay between the study of teaching / learning and the re-examination of physics--that enhance student learning.

First, it appears that teaching a topic forces an added level of reflection both upon the content and about an individual's own mastery of the subject. In support of this work, data from the class are being analyzed to observe the effects of teaching. Preliminary analyses of student performance indicate that students are more likely to master a subject conceptually if they teach the subject than if they cover textbook homework problems for the same amount of time. However, more data and analysis are required to make any definitive judgment.

A related line of research explores a critical link between student, content mastery and local context. Following the work of Cole (1996, 1998) current efforts address the relevance of researching student learning in context. In studying these environments and their implications for student learning, Cole argues that assessment of cognitive ability is contextually dependent -- that is, the further removed an experiment or study is from the domain of use/ application, the less applicable the result is to that domain. This same notion has been reported in a somewhat different form in physics. Studies, such as those using the Force Concept Inventory (FCI), report that while students may perform well in a traditional physics course, a course in which they have managed to master formulae and various mathematical procedures, the students miss the broader setting and conceptual basis for the discipline of physics (Redish, 2003). The first step in this exploration has been to develop a model of how context may be brought into the present research discussions on student learning of physics content (Finkelstein 2001; 2004).

Lastly, a rich area for investigation is the effect such a course has on students crossing disciplinary boundaries, and in particular, whether education can become a legitimate pursuit for physicists. There is evidence that this course helped students cross disciplinary boundaries. Of the six undergraduate physics majors in this study, four enrolled in teacher education programs. Three of these four enrolled in the university's teacher education program (tripling the annual enrollment of physics majors). A fifth student took a year abroad to teach and ultimately returned to enroll in an education program. Of the four graduate students enrolled, one took a post to in the physics department to direct the undergraduate laboratories. Another two were active participants in the American Association of Physics Teachers-sponsored graduate training program, Preparing Future Physics Faculty (PFPF), offered by the department. A more in-depth longitudinal study would be required to observe the longer term and broader impacts of the course. However, these preliminary signs indicate that by presenting the opportunity to explore and seriously consider education as a pursuit within the physics department, students begin to do just that.

*D. Institutional response --- Can this activity survive?*

The institutional response to this new activity system has not been a simple or monolithic process, and is worthy of a detailed research study in its own right; however, the program has initially met with some success. In Seymour Sarason's framework, this coordinated set of activities constitutes a new *setting*, or a new and sustained relationship between individuals (Sarason 1989; 1997). Sarason contends that program success or failure depend critically upon two factors: the initial structure of the program and the adaptation of that structure to local conditions.

After extensive setup (and lobbying of the department and other institutional entities), the course on Teaching and Learning Physics was incorporated into the institutional structure. The Department of Physics has adopted the course as an upper division restricted elective in the sequence of classes required for a bachelor's degree. The Teacher Education Program offers the class as part of its certification program. Additionally, this class is the first course to be cross- listed between the physics department and the Teacher Education Program.  In establishing and maintaining this course, an inter-disciplinary team has gathered to critique and help shape the course. Following the establishment of the course, the Director of Teacher Education Preparation and the Vice-Chair of Physics engaged in collaborative discussions surrounding the development of a new undergraduate physics and education major.  While the new major was not constructed, the course continues to the present, some several years later. As Sarason suggests, the creation of a new setting, one that often begins with multidisciplinary work, necessarily affects the local, disciplinary, and interdisciplinary cultures in which the new setting is created (Sarason 1989).

However, as Sarason goes further noting that these systems all-too-often depend upon a single individual and hence may not be able to adapt when that individual leaves (Sarason 1997). The course on Teaching and Learning Physics was offered by the author three times before leaving the institution.  Each of the first three sessions was offered voluntarily (that is, with funding support coming from a National Science Foundation Fellowship rather than the department).  With an eye to handing the course over, a second faculty member participated in the course the second time it was offered.  Currently, course is under the supervision of the new instructor and been offered twice, with the support of the department.  There is some indication that given the success of the course, it may continue; however, the difficulty of relying upon a single individual's efforts and advocacy remains; when the current instructor retires or moves on, it is not clear that the course will continue.

Beyond the local or micro-level, the course is part of a larger activity system of community partnership.  As described, fieldwork is an integral component of the course, and as such, has required the development and strengthening of ties with community partners.  The community agencies which host student-teachers from the university course, such as local schools and Boys and Girls Clubs, have indicated great interest in the continuation of collaborative efforts.  The community partners greatly value the added human resources of student-experts who participate in local activities, and in several cases used these added resources to develop new educational programs. Without the involvement of the undergraduates in the outreach process, two of the four community-based programs would not have operated.  Meanwhile the community-based programs serve as necessary resources for the university students and researchers who use these environments as laboratories for studying pre-college student learning.  In this way, it is not simply a matter of the university delivering outreach and programming, but rather a collaborative arrangement whereby both partners develop and benefit from the interaction.  Community-university partnership programs, using this model, continue to expand both in size and scope (into more schools and at more educational levels).

An unexpected outcome of offering the course was newfound collaboration with one of the local two-year colleges.  The chair of the local community college physics department participated in the class on Teaching and Learning Physics the first time it was offered.  Through the course and following years, members from both the community college system and the university system explored mechanisms to increase transfer rates from the two-year college into the university, and in particular, into physics. One project that stemmed from these discussions was the augmentation of the graduate training program, Preparing Future Physics Faculty (PFPF), with the opportunity for graduate students to teach students at the community college. Using materials from the course on Teaching and Learning Physics, graduate students developed and offered a new course to community college students. The first two offerings of the course were successful at providing graduate students valuable and authentic teaching experiences, exposing students at the two year college to physics and the university culture, and increasing the transfer of students from the community college to the university. Once again, however, whether these ties will be sustained with the absence of key individuals is unclear. One of the graduates of the PFPF program continues talks with the community college; however, the joint program has not been offered since the author left the university.

**V. Conclusions.**

The presented model for coordinating physics education, research, and community partnerships may be adopted more broadly within the (science) education community by substituting different content. There is nothing particular to physics, nor undergraduates in this model. The domain of examination could equally well have been Newtonian mechanics, or physical chemistry. The outcomes would be similar: increased student interest and ability in the science domain, increased attention to and interest in teaching and education, and development of community partnerships. Furthermore, the activity system provides a rich opportunity for science education research that is tightly coupled with and informed by educational reforms. As institutions of higher education begin to develop programs of discipline-based education research *within* the sciences (science departments, and in particular physics departments are hiring an increasing number of faculty into new lines physics education research / reform), this type of activity system provides an avenue to leverage local interest in reform, research, and community partnership. Because such a system addresses the multiple motives of physics, education, and outreach, the hope is that each domain would support the activity and would develop an authentic interest in sustaining a coordinated program. Of course, such change is local and depends simultaneously upon fertile ground (local or bottom-up support) and healthy conditions for growth (top-down support).

**References**

APS (1999). American Physical Society. Statement 99.2: RESEARCH IN PHYSICS EDUCATION, adopted May 21, 1999. http://www.aps.org/statements/99.2.html

Astin, A.W., Vogelgesang, L. J., Ikeda, E.K., and Yee, J A. (2000). *How Service Learning Affects Students*. Los Angeles, CA: Higher Education Research Institute, University of California, Los Angeles. Available at: http://www.gseis.ucla.edu/slc/rhowas.html

Brown, A.L. (1992). Design Experiments: Theoretical and Methodological Challenges in Creating Complex Interventions in Classroom Settings, *Journal of the Learning Sciences*, **2**(2), 141-178.

Brown, A.L. and Campione, J.C. (1990). Communities of Learning and Thinking, or A Context by Any Other Name, *Human Development*, **21**, 108-126.